\PassOptionsToPackage{colorlinks=true, linkcolor=mydarkblue, citecolor=mydarkblue, filecolor=mydarkblue, urlcolor=mydarkblue}{hyperref}
\documentclass[conference, twocolumn]{IEEEtran}
\usepackage[utf8]{inputenc}
\usepackage{fontspec}

\usepackage{graphicx}
\usepackage{doi}
\usepackage{array}
\usepackage{xeCJK}
\usepackage{listings}
\usepackage{subcaption}
\usepackage[table,xcdraw]{xcolor}
\usepackage{amsmath}
\usepackage{bbm}
\usepackage{url}
\usepackage{pifont}
\usepackage{acronym}
\usepackage{amssymb}
\definecolor{mydarkblue}{rgb}{0,0.08,0.65}
\usepackage[colorlinks=true,
    linkcolor=mydarkblue,
    citecolor=mydarkblue,
    filecolor=mydarkblue,
    urlcolor=mydarkblue]{hyperref}
\usepackage{cleveref}
\usepackage{soul}
\usepackage[shortlabels]{enumitem}
\usepackage{color}
\usepackage{tikz}
\usepackage{balance}
\usepackage{multirow}
\usepackage{comment}
\usepackage{wrapfig,booktabs}
\usepackage[frozencache,cachedir=.]{minted}

\usepackage[symbol]{footmisc}
\usepackage{tablefootnote}
\usepackage{tabularx}
\usepackage{placeins}
\usepackage{algorithm}
\usepackage{algpseudocode}
\usepackage{ragged2e}
\usepackage{cuted}
\usepackage{float}
\usepackage{caption}

\usepackage{multicol}

\usepackage{dblfloatfix}
\usepackage{natbib}

\renewcommand{\arraystretch}{1.2}

\definecolor{codegreen}{rgb}{0,0.6,0}
\definecolor{codegray}{rgb}{0.5,0.5,0.5}
\definecolor{codepurple}{rgb}{0.58,0,0.82}
\definecolor{backcolour}{rgb}{0.95,0.95,0.92}

\makeatletter
\def\blfootnote{\xdef\@thefnmark{}\@footnotetext}
\makeatother

\lstdefinestyle{mystyle}{
  backgroundcolor=\color{backcolour},   commentstyle=\color{codegreen},
  keywordstyle=\color{magenta},
  numberstyle=\tiny\color{codegray},
  stringstyle=\color{codepurple},
  basicstyle=\ttfamily\footnotesize,
  breakatwhitespace=false,
  breaklines=true,
  captionpos=b,
  keepspaces=true,
  numbers=left,
  numbersep=5pt,
  showspaces=false,
  showstringspaces=false,
  showtabs=false,
  tabsize=2,
}

\AtBeginDocument{%
  \providecommand\BibTeX{{%
    \normalfont B\kern-0.5em{\scshape i\kern-0.25em b}\kern-0.8em\TeX}}}

\IEEEoverridecommandlockouts

\acrodef{TTS}{Text To Speech}
\acrodef{MoE}{Mixture of Experts}
\acrodef{RVQ}{Residual Vector Quantization}
\acrodef{DAC}{Descript Audio Codec}
\acrodef{ASR}{Automatic Speech Recognition}
\acrodef{VAD}{Voice Activity Detection}
\acrodef{WER}{Word Error Rate}
\acrodef{MLP}{Multi Layer Perceptron}
\acrodef{CCA}{Compressed Convolutional Attention}
\acrodef{GQA}{Grouped Query Attention}
\acrodef{EDA}{Exponential Depth Averaging}
\acrodef{UTMOS}{UTokyo-SaruLab Mean Opinion Score}
\acrodef{MHA}{Multi-Head Attention}
\acrodef{GELU}{Gaussian Error Linear Units}
\acrodef{RoPE}{Rotary Positional Embeddings}
\acrodef{DS-WED}{Discretized Speech Weighted Edit Distance}
\acrodef{SNR}{Signal to Noise Ratio}
\acrodef{LDA}{Linear Discriminant Analysis}
\acrodef{TTSDS2}{Text-to-Speech Distribution Score 2}
\acrodef{G2P}{Grapheme to Phoneme}
\begin{document}

\title{ZONOS2 Technical Report}

\newcommand{\corr}{\textsuperscript{*}}

\author{
Gabriel Clark\corr, Sofian Mejjoute, Mohamed Osman, George Close, \\ Beren Millidge\corr
\\[0.5em]
\textbf{Zyphra}
\\
San Francisco, CA \\
\IEEEauthorblockA{\textsuperscript{*}Corresponding authors: \texttt{gabriel@zyphra.com}, \texttt{beren@zyphra.com}}
}

\maketitle

\setcounter{page}{1}

\begin{abstract}
We present ZONOS2 8B, our latest \ac{TTS} model, which achieves state-of-the-art naturalness, prosody, and voice cloning fidelity. We improve upon Zonos-v0.1 in scale, data, and training recipe. We scale the model from 1.6B to 8B total parameters (900M active) with a novel mixture-of-experts (MoE) backbone, improving inference latency and throughput. We expand our training corpus from 200K to over 6M hours using a new data processing pipeline, and we simplify our post-training and conditioning recipes to improve naturalness and voice cloning fidelity. We evaluate ZONOS2 8B on quality, speaker similarity, WER, and ZTTS1-Eval, our novel TTS benchmark, where it performs competitively with state-of-the-art systems while maintaining good streaming latency. We release our model weights and example inference code under an Apache 2.0 license on GitHub\footnote{https://github.com/Zyphra/ZONOS2/} and Hugging Face\footnote{https://huggingface.co/Zyphra/ZONOS2}, as well as the ZTTS1-Eval benchmark\footnote{https://github.com/Zyphra/ZTTS1-Eval}.
\end{abstract}

\begin{figure*}[t]
    \centering
    \includegraphics[width=\linewidth]{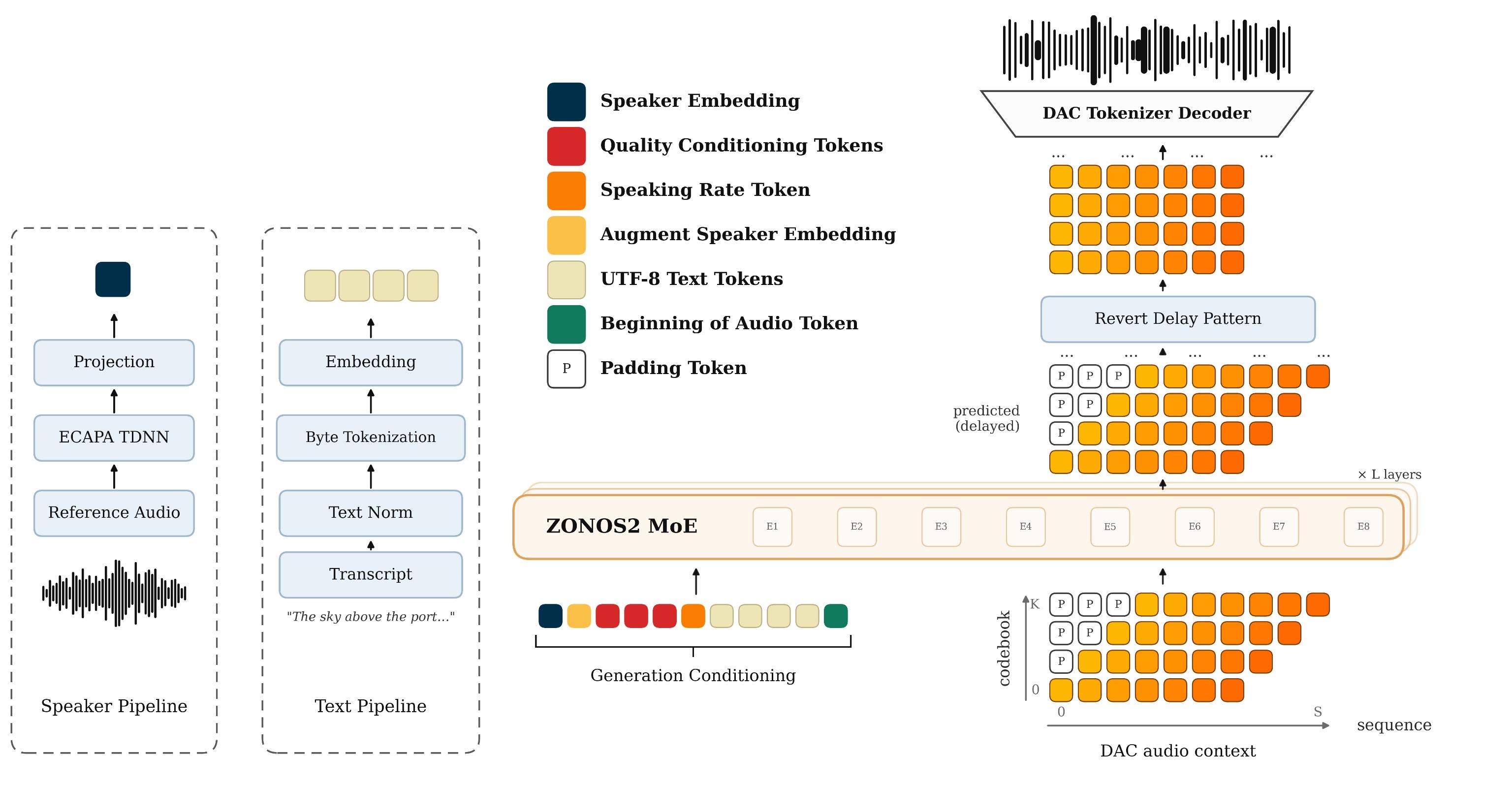}
    \caption{Overview of ZONOS2 inference pipeline, showing the text and conditioning inputs as well as the delay pattern approach to audio codec token generation.}
    \label{fig:overview}
\end{figure*}

\section{Introduction}
The core quality measures of a \ac{TTS} system are generation stability, naturalness, controllability, multilinguality, and inference performance. Recent works in this domain tend to focus on one of these properties to the detriment of the others. For instance, some systems produce high-quality output but offer limited control, while inference-efficient systems support only one or a few languages. With ZONOS2, our aim is to design a \ac{TTS} system that is strong in every aspect, focusing especially on naturalness and voice cloning fidelity.

The ZONOS2 architecture uses a decoder-only transformer backbone which takes text and audio tokens as input and outputs audio tokens, where the audio tokens are represented using a \ac{RVQ}-based audio codec. Our architectural novelty stems from introducing \ac{MoE} architectures into open-source TTS, which we use to scale the model's parameter count and capability while maintaining inference speeds necessary for real-time audio streaming. We combine architecture advancements from recent LLMs, including the ZAYA1-8B~\citep{washbourne2026zaya18btechnicalreport} router design. To maximize the quality of the generated audio, a high fidelity and high sample rate codec~\citep{kumar2023highfidelityaudiocompressionimproved} \ac{DAC} is used.
To provide controllability, we implement several user-facing conditioning options, including high accuracy voice cloning from prompt audio, speaking-rate selection, and an optional `Quality Mode'. Our voice cloning approach is zero-shot, does not require speaker labels during pretraining, supports unbounded speaker utterance lengths, and does not require transcription of the clone target audio at inference time. For seamless multilingual performance and robustness to phonemization errors, we adopt raw byte tokenization for the text input. We increased the linguistic diversity of our training set, focusing heavily on broadening support for European and Asian languages. To achieve low latency, we use a \textit{delay pattern}~\citep{copet2024simplecontrollablemusicgeneration} approach to codebook generation which generates multiple codebook tokens in parallel by overlapping the generation of different codebooks of different timesteps. Moreover, our MoE architecture significantly improves inference latency due to its small forward pass footprint. 

Finally, motivated by the limitations of existing \ac{TTS} evaluation benchmarks, we also introduce a new benchmark, ZTTS1-Eval. The most widely adopted \ac{TTS} benchmark, Seed-TTS-Eval~\citep{anastassiou2024seedttsfamilyhighqualityversatile}, covers only English and Chinese, scores intelligibility and speaker similarity with now-dated models, and uses only read speech. Later efforts such as CV3-Eval~\citep{du2025cosyvoice} broaden language and task coverage but keep the same scoring stack and a largely prepared-speech setting. As modern systems now approach human intelligibility, \ac{WER} alone no longer separates the strongest systems, and no widely used zero-shot benchmark measures whether generated speech reproduces the prosody of real speech. ZTTS1-Eval addresses this. It pairs a clean read-speech set of 9 languages with an in-the-wild spontaneous set of 17 languages, scores content, speaker, and quality with current models (Qwen3-ASR~\citep{Qwen3-ASR}, ReDimNet~\citep{Yakovlev_2024}, and MSR-UTMOS~\citep{nishikawa2025multisamplingfrequencynaturalnessmosprediction}), and adds a distributional prosody and generation-diversity layer (TTSDS2~\citep{minixhofer2025ttsds2} and \ac{DS-WED}~\citep{yang2026dswed}). We describe the full details of our ZTTS1-Eval benchmark in Section~\ref{sect:ztts1}.

Overall, our key contributions are:
\begin{itemize}
    \item We release ZONOS2, an open-source, permissively licensed TTS model with high-fidelity natural-sounding voice cloning.
    \item We pioneer the use of \ac{MoE} models in the open-source TTS space, building upon the ZAYA1 architecture~\citep{washbourne2026zaya18btechnicalreport}.
    \item We introduce a new method for post-training high fidelity voice conditioning without labeled speaker pairs.
    \item We detail our novel data curation pipeline for web-scale data, which enables several inference conditioning settings and multilingual generation.
    \item We release a new \ac{TTS} benchmark, ZTTS1-Eval, spanning 9 read and 17 in-the-wild languages with measures for prosody, speaker similarity, and generation quality.
\end{itemize}

The remainder of this technical report is organized as follows. In Section~\ref{sect:overview}, we give an overview of the ZONOS2 model. Section~\ref{sect:data} describes the training data pipeline. Section~\ref{sect:training} describes the model training setup. Section~\ref{sect:ztts1} details our newly proposed benchmark. Section~\ref{sect:results} presents the performance of the ZONOS2 model relative to other state-of-the-art \ac{TTS} systems on several \ac{TTS} evaluation sets, including ZTTS1-Eval. Section~\ref{sect:discuss} discusses the challenges we faced during the training of ZONOS2 and our future research directions.

\section{Model Overview}\label{sect:overview}

ZONOS2 is built on a transformer MoE backbone with 900M active parameters and 8B total parameters. It uses standard autoregressive language modeling of \ac{RVQ}-based audio tokens. Figure~\ref{fig:overview} provides an inference overview of ZONOS2.

\subsection{Audio Tokenization}
To maximize the fidelity of the generated audio, ZONOS2 uses \ac{DAC}~\citep{kumar2023highfidelityaudiocompressionimproved} as the neural audio tokenizer. \ac{DAC} follows an autoencoder structure, where the encoder maps raw waveforms $x$ to latent representations which are quantized into $N$ codebooks using a \ac{RVQ}~\citep{zeghidour2021soundstreamendtoendneuralaudio} strategy. The decoder then outputs the reconstructed audio from the quantized codebook sequence. 
ZONOS2 models these quantized codebooks using a \textit{delay pattern} which is depicted in Figure~\ref{fig:overview} and described here. Let $X[t,j]$ denote the aligned DAC token for audio frame $t$ and codebook $j$, where $0 \le j < N$ and $N = 9$ in this work. The delay pattern is built by a shearing operation:
\begin{equation}
    Y[t, j] =
    \begin{cases}
        X[t-j, j] & \text{if } t \ge j, \\
        p & \text{otherwise},
    \end{cases}
\end{equation}
where $p$ is the ID of a special audio padding token. Thus, codebook $j$ is delayed by $j$ frames. The delay pattern turns the within-frame dependency among \ac{RVQ} codebooks into an autoregressive dependency over sequence positions: the token for codebook $j+1$ at audio frame $t$ is generated immediately after the token for codebook $j$ at the same frame. As a result, the model can predict each codebook token while conditioning on preceding codebooks for that frame, rather than treating the codebooks as conditionally independent. Before DAC decoding, the shearing operation is inverted:
\begin{equation}
    \hat{X}[t,j] = Y[t+j,j].
\end{equation}
As a result, streaming decoding requires a lookahead buffer of $N-1$ generated frames before all codebooks for an aligned audio frame are available and the tokens from that frame can be passed to the \ac{DAC} decoder. 

\subsection{Text Tokenization}
ZONOS2 tokenizes text with a byte-level tokenizer. An input string is encoded as a sequence of UTF-8 bytes $(b_1, \ldots, b_L)$, with each $b_i \in [0, 255]$. This representation lets the model consume arbitrary Unicode input without language-specific \ac{G2P} conversion or an out-of-vocabulary mechanism. Our previous system, Zonos-v0.1, operated on phonemes since phonemization injects a strong inductive bias that accelerates convergence early in training. However, this inductive bias originates from a G2P phonemization pipeline for which coverage and accuracy deteriorate for lower-resource languages, for code-switched (i.e., containing words from multiple languages) utterances, and for rare or technical vocabulary. With ZONOS2 and our increased data scale, we found that removing the crutch of this inductive bias proved helpful overall, especially improving the model's multilingual capabilities and general robustness. 

Table~\ref{tab:g2p-errors} shows three representative \ac{G2P} failures with different failure modes. In the code-switched Chinese-English example, the Chinese span appears in text tagged as English. Rather than invoking a Mandarin-aware frontend with segmentation, tone handling, and lexicon lookup, the pipeline falls back to generic character labels before processing the English words. In the second example, an overgeneralized substring or morphology rule incorrectly matches the sequence \textit{retro} inside \textit{alpharetrovirus}, producing the spurious form \textit{retroretrovirus}. In the Spanish example, the proper noun \textit{Satoshi} is processed under the Spanish language tag and constrained to the Spanish phoneme inventory used by the \ac{G2P} phonemizer. Because that inventory lacks a fricative corresponding to the romanized sequence \textit{sh} in \textit{shi}, the sequence is mapped to \textit{s}, yielding \textit{Satosi} and losing the intended source-pronunciation cue. All three failures are silent. The downstream model receives a corrupted transcription with no indication that preprocessing has failed, making this kind of error especially harmful in low-resource and code-switched settings.

\begin{table}[h]
\centering
\begin{tabular}{@{}l|l@{}}
\toprule
Input & G2P pipeline output \\
\midrule
en:电脑 is computer & en:Chinese letter Chinese letter is computer \\
en:alpharetrovirus & en:retroretrovirus \\
es:Satoshi & es:Satosi \\
\bottomrule
\end{tabular}
\caption{Representative silent failures in the \ac{G2P} preprocessing pipeline. The first row shows a Chinese span processed under an English tag; the second shows an overmatched substring or morphology rule; the third shows loss of source-pronunciation information when a proper noun is constrained to the Spanish \ac{G2P} inventory.}
\label{tab:g2p-errors}
\end{table}

Early experimentation revealed that the value of the inductive bias that phonemization provides diminishes with scale. As data and model size are increased, the model learns the mapping from raw bytes to pronunciation on its own. A byte-level variant first matches and then surpasses its phoneme-based counterpart while avoiding the failure modes of \ac{G2P} entirely. Bytes are adopted as the simplest representation that is both fully general across languages and free of an error-prone preprocessing stage. During inference, text normalization is applied in several cases to allow for consistent pronunciation of equations, numbers, addresses, etc.

\subsection{Speaker Embeddings}
Speaker identity information is necessary for zero-shot speaker cloning. For ZONOS2, we condition on ECAPA-TDNN speaker embeddings~\citep{desplanques2020ecapatdnn}, using the pre-trained module of \citet{hu2026qwen3ttstechnicalreport} to extract a $2048$-dimensional embedding $\mathbf{e}_\mathrm{x}$ from a given utterance $x$.

Conditioning on this single embedding vector, rather than on the reference waveform or its token sequence, is attractive for two reasons. First, the speaker representation occupies a single position at the start of the sequence, so it adds negligible cost to the prefix and does not grow the context the decoder must attend over, regardless of how long the reference recording is. This avoids the problem of long clone target utterances consuming context which would otherwise be used for generation, as well as reliance on a transcription of the speech in the target clone utterance. Second, the embedding is high-bandwidth; a single $2048$-dimensional vector can capture nearly all of the desired speaker characteristics, so little identity information is lost relative to conditioning on the full utterance.

Increasing the speaker embedding bandwidth means that the embedding also encodes a great deal of undesired information. The duration of the reference utterance, non-speech acoustic properties such as background noise and recording conditions, the exact words spoken in the prompt, and the placement of pauses are all present in $\mathbf{e}_\mathrm{x}$. Importantly, none of these should be carried into the generated speech which we expect to be different along these axes. To reduce this kind of unintentional leakage, the embedding is projected through an \ac{LDA} transform to a $1024$-dimensional vector $\hat{\mathbf{e}}_\mathrm{x}$.
Because the \ac{LDA} is estimated from speaker-labeled data to maximize between-speaker variance relative to within-speaker variance, it retains the directions that separate one speaker from another while attenuating the factors that vary across different recordings of the same speaker such as duration, noise, lexical content, and pause structure noted above.

To condition the autoregressive \ac{TTS} model on this representation, a learned projection of the speaker embedding is placed at the beginning of the sequence:
\begin{equation}
    h_{\mathrm{spk}} = W_{\mathrm{spk}} \hat{\mathbf{e}}_\mathrm{x} + b_{\mathrm{spk}},
\end{equation}
where $W_{\mathrm{spk}} \in \mathbb{R}^{d_{\mathrm{model}} \times 1024}$ maps the speaker embedding into the transformer hidden dimension.

Beyond discarding nuisance factors, the \ac{LDA} projection helps mitigate a core problem with the speaker embeddings. There is a tradeoff between how much information an embedding carries about its source utterance and how long the model can be trained on embeddings drawn from that utterance before overfitting starts to occur. Because the embedding is computed from the ground-truth recording of the target utterance, any utterance-specific information it leaks, such as lexical content and pause timing, offers the model a shortcut. It can reduce its training loss by reading these specifics out of the speaker embedding instead of learning a solution that generalizes to text, emotion, or pauses not present in the reference clip. The more information about the target the embedding exposes, the sooner this shortcut comes to dominate and the fewer useful training steps remain. 
Without the LDA transformation, we found we could not take enough training steps with the speaker embedding to learn high-quality voice cloning before overfitting occurred. The \ac{LDA} projection is necessary to extend the usable training horizon of the embedding.

Even with the \ac{LDA} projection, the extent to which the model can be safely trained using embeddings extracted from the target utterance remains limited. The training horizon is extended with a two-phase scheme for speaker conditioning, described in Section~\ref{conditioning-fine-tuning}, which is ultimately what yielded the final ZONOS2 model.

\subsection{Speaking-Rate Conditioning}
To enable finer-grained control over speech generation, speaking-rate conditioning is introduced. For each training utterance, symbols, annotations, whitespace, and punctuation are first stripped from the text, and the speaking rate is computed as the number of UTF-8 bytes in the remaining text divided by the utterance duration in seconds. This rate is quantized into a discrete bucket and a token representing the bucket is prepended to the utterance's text tokens in the packed sequence.

\subsection{Quality Conditioning}
During initial testing, the model was found to be highly sensitive to the acoustic content of the speaker-cloning audio which degraded the quality of generations from low quality speaker samples. To produce consistently high-quality generations regardless of the acoustic properties of the clone audio, two forms of quality conditioning are introduced.

First, during training, various acoustic augmentations are applied to the clone audio before the speaker embedding is computed. These augmentations include mixing in background noise or music, applying audio-codec compression, and simulating environmental reverberation. The target \ac{DAC} tokens are computed from the clean, un-augmented audio, so that the model learns to be robust to these augmentations in the clone audio. This augmentation is applied with probability $\alpha_{\mathrm{AUG}}$ during annealing, and only to utterances whose metadata indicates the source audio is clean. When applied, an additional `Augmented Embedding' token is appended to the sequence.

Second, additional conditioning information, such as generation bandwidth, volume, leading and trailing silent frames, and estimated \ac{SNR}, is also introduced as synthetic text tokens, following the same approach used for the speaking-rate conditioning described above. This enables fine-grained control over the acoustic properties of the generated audio at inference time. During the final annealing stage, a high-quality subset of the training corpora is paired with a `Quality Mode' token. At inference, this allows the user to select high intelligibility at the cost of speaker-cloning capability.

\subsection{MoE Transformer}\label{sec:transformer_layers}
\begin{figure}
    \centering
    \includegraphics[width=.9\linewidth]{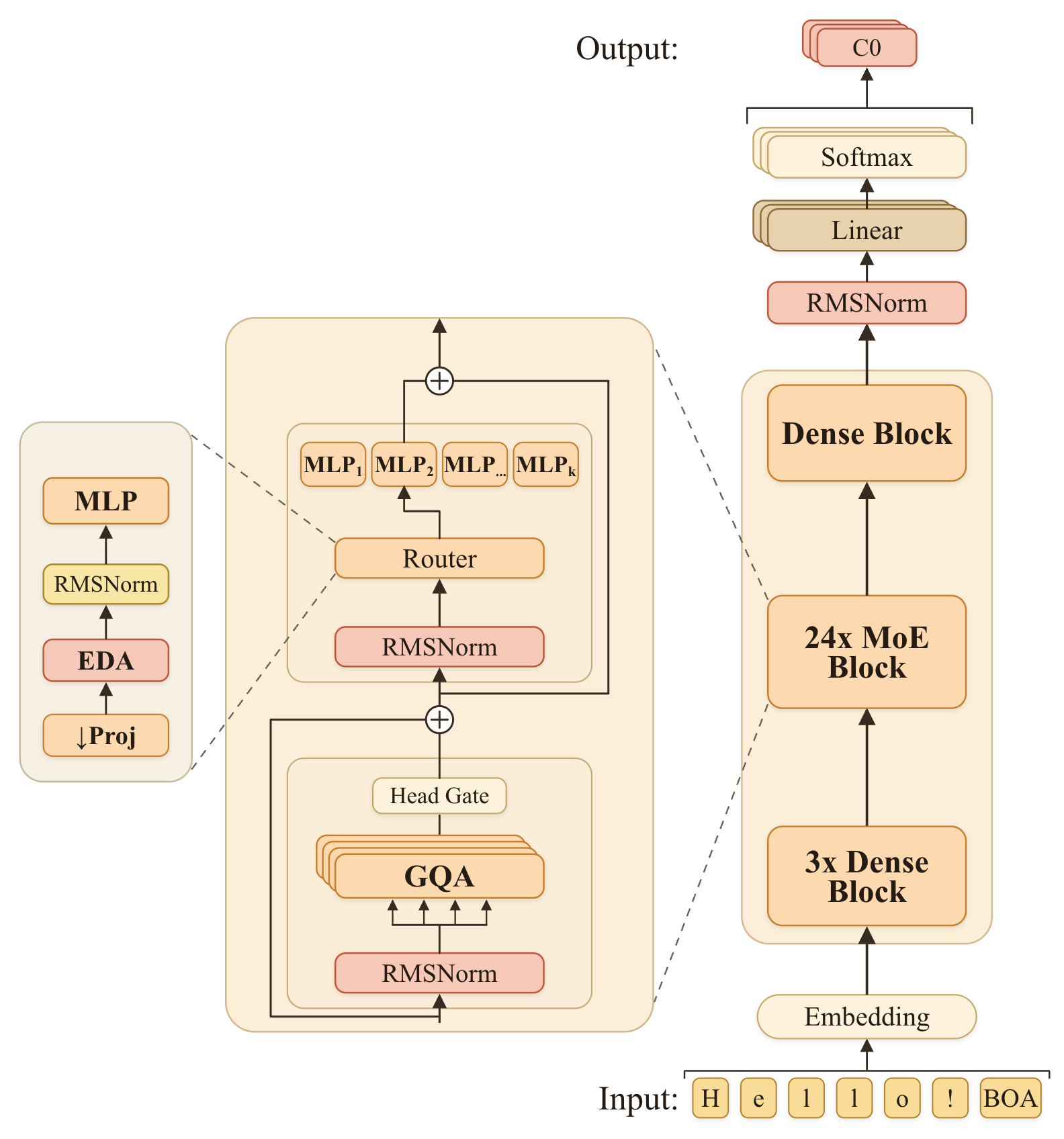}
    \caption{Schematic of the ZONOS2 transformer MoE architecture.}
    \label{fig:moe_transformer}
\end{figure}

The backbone is a $28$-layer transformer with hidden dimension $2048$ (Figure~\ref{fig:moe_transformer}). The first three layers and the final layer are dense; the rest follow the \ac{MoE} transformer and use top-$1$ routing, except the final routed layer, which uses top-$2$. The router design closely follows \citep{washbourne2026zaya18btechnicalreport}, and attention uses \ac{GQA}~\citep{ainslie2023gqatraininggeneralizedmultiquery} with Qwen gating~\citep{qwen-gating} in the headwise location. The input representation is formed by embedding each delayed audio-codebook frame with its own embedding table and summing the resulting vectors; when speaker conditioning is used, the projected embedding is placed at the start of this sequence, which is then normalized with $\mathrm{RMSNorm}$~\citep{zhang2019rootmeansquarelayer} before being passed to the transformer stack.

The router's network structure is shown on the left of Figure~\ref{fig:moe_transformer}. A down projection feeds an \ac{EDA} stage~\citep{pagliardini2024denseformerenhancinginformationflow,washbourne2026zaya18btechnicalreport}, which blends the previous layer's router states with the current layer's via a learned scaling parameter. The result passes through $\mathrm{RMSNorm}$ and a three-layer \ac{MLP} with \ac{GELU} activations~\citep{hendrycks2023gaussianerrorlinearunits} to produce the final router scores. For a full configuration breakdown see Table~\ref{tab:zonos_config}; for full model details see Section~\ref{sect:model_details}.

\subsection{Training Objective}\label{sec:training_objective}
The model is trained as a standard autoregressive conditional language model, outputting \ac{DAC} tokens. Let $s$ denote the complete packed training sequence, consisting of the optional speaker-conditioning and speaking-rate frames, the byte-tokenized text prompt, and the delayed audio-token frames. Given the causal hidden state $h_t=f_\theta(s_{<  t})$, the model predicts the next delayed audio-codebook frame:
\begin{equation}    
p_\theta(Y[t+1,n]=v \mid s_{< t})
=
\operatorname{softmax}(\tilde{\ell}_{t,n})_v.
\end{equation}

For numerical stability, logits are soft-capped before the softmax~\citep{gemmateam2024gemma2improvingopen}:
\[
\tilde{\ell}_{t,j}
=
\tau \tanh\left(\frac{\ell_{t,j}}{\tau}\right),
\]
with $\tau = 15$ in our training runs. The per-codebook predictive distribution
is then
\begin{equation}
p_\theta(Y_{t,j}=v \mid s_{< t})=
\operatorname{softmax}(\tilde{\ell}_{t,j})_v.
\end{equation}

The main training loss is the masked negative log-likelihood over non-padding
audio targets:
\begin{equation}    
\mathcal{L}_{\mathrm{NLL}}
=
-\frac{1}{M_{\mathrm{aud}}}\sum_{t,j}
m_{t,j}
\log p_\theta(Y[t+1,j] \mid s_{< t}),
\end{equation}
where $m_{t,j}=1$ only when the target $y_{t,j}$ is not $p$, and
$M_{\mathrm{aud}}=\sum_{t,j} m_{t,j}$. Text tokens and optional speaker-conditioning, speaking-rate and quality positions are therefore used as context but are not reconstructed as targets.

For mixture-of-experts layers, a router balancing objective is used during training. Let $u_{\ell,e}$ be the empirical fraction of routed tokens assigned to expert $e$ in MoE layer $\ell$, and let $\bar{u}_e = 1/E$ be the uniform expert usage for $E$ experts~\citep{wang2024auxiliarylossfreeloadbalancingstrategy}. Each MoE layer maintains a zero-mean balancing
bias vector $b_\ell$. The auxiliary loss is
\begin{equation}    
\mathcal{L}_{\mathrm{bal}}
=
\sum_{\ell \in \mathcal{M}}
b_\ell^\top
\operatorname{sg}(u_\ell - \bar{u}),
\end{equation}

where $\operatorname{sg}(\cdot)$ denotes stop-gradient. Minimizing this term
decreases the routing bias for overused experts and increases it for underused
experts. A separate AdamW~\citep{loshchilov2018decoupled} optimizer is used to learn the bias vectors. 

The total objective minimized during training is simply the sum of these terms:
\begin{equation}    
\mathcal{L}
=
\mathcal{L}_{\mathrm{NLL}}
+
\mathcal{L}_{\mathrm{bal}}.
\end{equation}

\begin{figure}[t]
    \centering
    \includegraphics[width=0.9\linewidth]{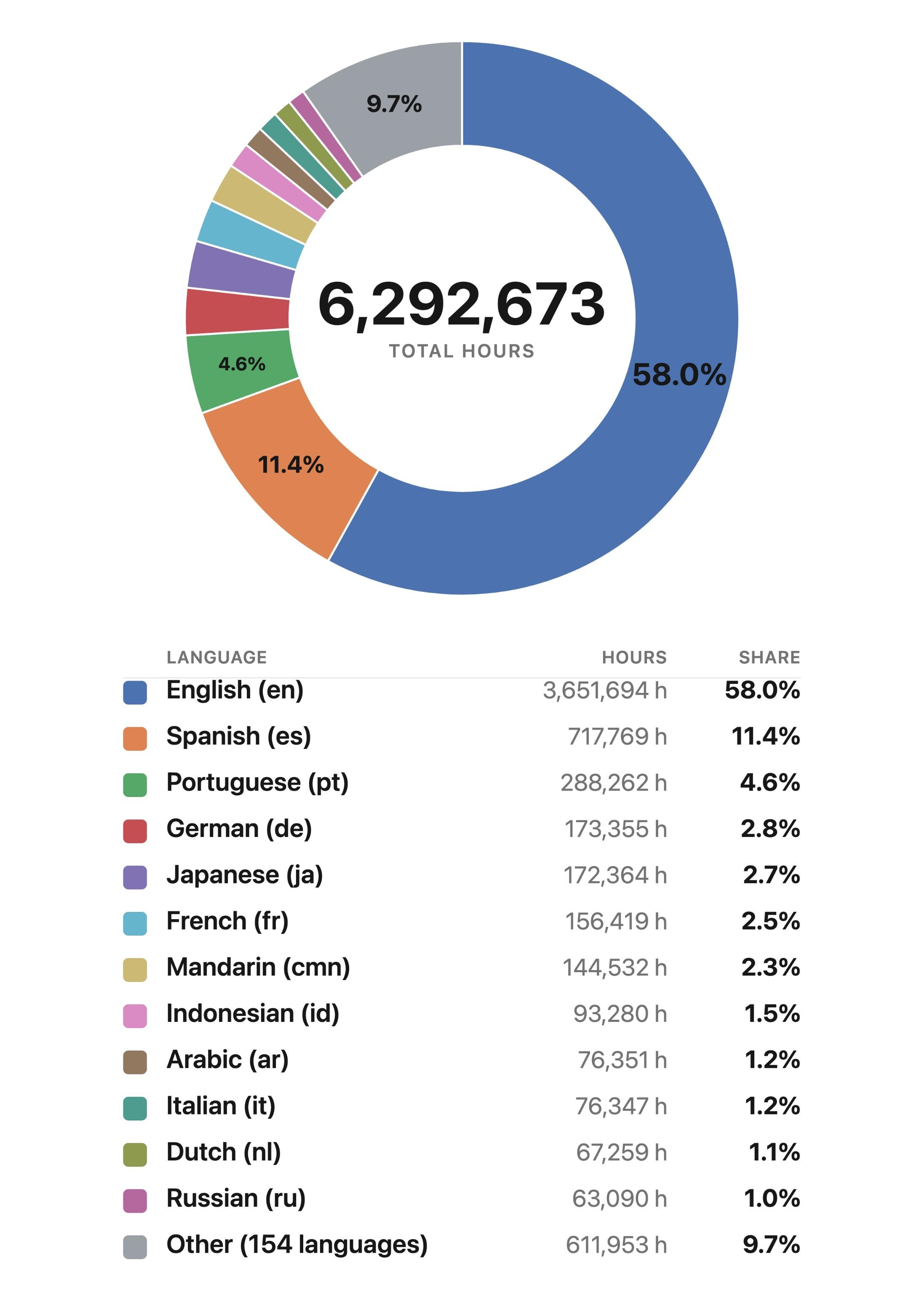}
    \caption{A breakdown of the training dataset for ZONOS2 by language.}
    \label{fig:hours_pie}
\end{figure}

\section{Data Pipeline}\label{sect:data}

\begin{figure*}[!t]
    \centering
    \includegraphics[width=0.9\linewidth]{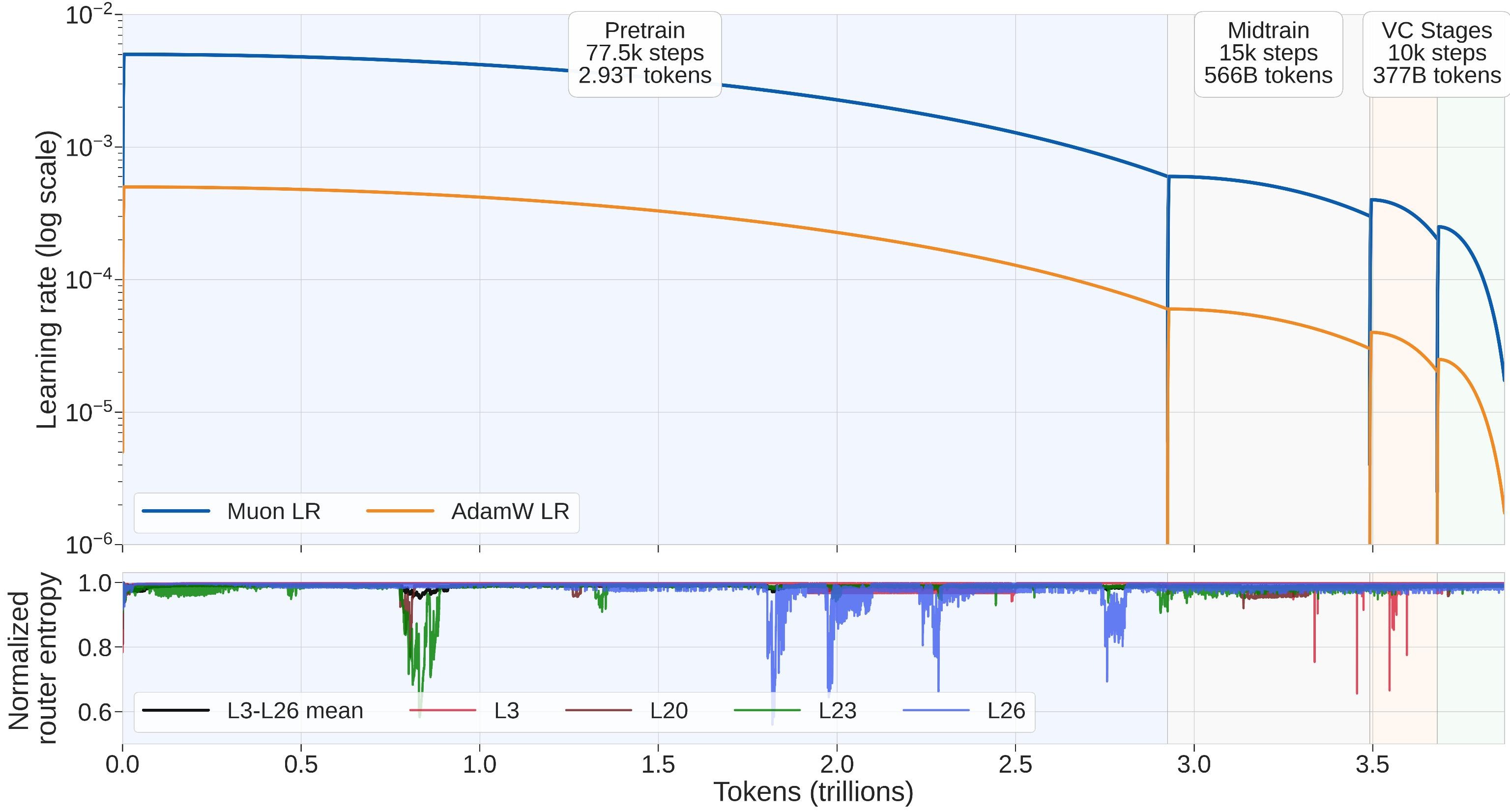}
    \caption{Learning rate and \ac{MoE} router entropy over each stage of training.}
    \label{fig:lr_plot}
\end{figure*}

The key to high-quality \ac{TTS} training data is strong alignment between audio and its transcription. To this end, we implemented a two-stage data processing pipeline. For the first stage, we applied a \ac{VAD} system to all raw audio, producing utterance-level segments. In the second stage, we used multiple \ac{ASR} systems to independently transcribe each utterance~\citep{koluguri2025granaryspeechrecognitiontranslation}.

This ensemble approach to transcript generation has several advantages. During training, we measured inter-ASR-system agreement via the pairwise \ac{WER} between ASR transcripts, where a lower \ac{WER} indicates closer agreement. We discarded utterances where the pairwise error exceeded a minimum threshold, removing low-quality samples on which the ASR systems disagree. We then adjusted the threshold to the specific needs of different training stages. It was set low during pre-training, where data variety matters most, and raised during annealing where the character of the final model is formed. The ensemble also allowed different transcripts to be selected as the input text for the same utterance across training, preventing the \ac{TTS} model from overfitting to a particular \emph{style} of transcription.

Our full dataset combines public speech corpora, podcasts, public-domain audiobooks, conversational datasets, multilingual web-scale speech data, and expressive or character-driven voice datasets. Data loading was configured so that specific sub-datasets can be up-weighted during training. In Figure~\ref{fig:hours_pie}, we break down our entire $6.2$-million-hour dataset by language. While English forms the majority of the training data, the final model generalizes well across all languages tested, including those which represent only a fraction of the overall training data.

\section{Training Overview}\label{sect:training}

The training process consisted of four stages. The first was a large-scale pre-training stage without conditioning. This was followed by a shorter mid-training stage with stricter transcript agreement filtering and sub-dataset selection. An initial annealing stage introduced conditioning signals like the speaker embedding, speaking-rate tokens, and quality conditioning. The first annealing stage also embedded only segments of the target audio with the speaker embedding and masked the loss for those segments to prevent causal leakage. This was followed by a final annealing stage in which the `Quality Mode' conditioning was introduced, speaker embeddings cover the entire target segment, and loss masking was dropped for the embedded speaker segment. Figure~\ref{fig:lr_plot} shows the overall learning rate as well as \ac{MoE} router entropy for the most unstable layers of the model during each stage of training.

\subsection{Pre-training}
The model was pre-trained for $77{,}500$ optimizer steps or 2.9T tokens. This stage used the base \ac{TTS} objective without speaker embeddings or quality conditioning. Text was tokenized as bytes, and audio was represented as discrete multi-codebook audio tokens, as described above.

\begin{table*}[ht]
\centering
\begin{tabular}{lcccc}
\hline
 & Seed-TTS-Eval & CV3-Eval & MiniMax-ML & \textbf{ZTTS1-Eval} \\
\hline
Languages        & 2                      & 9                      & 24                     & \textbf{up to 17} \\
Audio            & read                   & read/expressive        & read                   & \textbf{read and ITW spontaneous} \\
Duration         & 3h                     & 14h                    & --\textsuperscript{$\dagger$} & \textbf{16h} \\
Prosody / div.   & --                     & task specific          & --                     & \textbf{TTSDS2 + DS-WED} \\
ASR scorer       & Whisper-L / Paraformer & Whisper-L / Paraformer & Whisper-L / Paraformer & \textbf{Qwen3-ASR} \\
Speaker scorer   & WavLM                  & ERes2Net               & WavLM                  & \textbf{ReDimNet} \\
Quality scorer   & --                     & DNSMOS                 & --                     & \textbf{MSR-UTMOS} \\
\hline
\end{tabular}
\caption{Comparison of Seed-TTS-Eval, CV3-Eval, the MiniMax multilingual test set (MiniMax-ML), and the proposed ZTTS1-Eval. $\dagger$: MiniMax-ML contains 100 sentences and two reference utterances per language.}
\label{tab:tts-eval-comparison}
\end{table*}

We trained at a maximum sequence length of $6144$ frames and sequences were packed together with document masking to create global batches of $37.7$ million \ac{DAC} frames or $121.8$ hours of audio. Data sources were sampled with dataset-specific weights to increase the frequency of underrepresented and high quality audio types such as expressive speech, acted speech, and dialogue-style data relative to larger generic speech corpora.

We trained with the Muon optimizer~\citep{jordan6muon,moonshot-muon} with a base learning rate of $5 \times 10^{-4}$, a Muon learning rate of $5 \times 10^{-3}$, weight decay of $0.1$, gradient clipping at $0.5$, a $100$-step warmup, and cosine learning-rate decay.

During the pre-training stage, we found that the expert routing was highly unstable and the normalized entropy of several layers periodically collapsed to as low as $0.6$ and remained there without intervention. Normalized entropy for unbalanced layers is visible in Figure~\ref{fig:lr_plot}. To mitigate this instability, we set the first three layers and the final layer to be dense transformer blocks, and the last \ac{MoE} block to use top-$2$ routing. In addition, the balancing-bias and router learning rates were adjusted manually in response to the different types of expert collapse encountered during training. We found that MoE balancing on audio data is substantially harder than on text data, for reasons we do not fully understand. We speculate that this could be due to the intrinsic difficulty and noisiness of delayed DAC tokens, the unique challenge of aligning high frame rate audio with low frame rate text tokens, or other statistical properties of audio data.

\subsection{Mid-training}
This stage was largely the same as pre-training except that inter-\ac{ASR} system transcript agreement was increased compared to pre-training. Mid-training was run for $15{,}000$ optimizer steps or approximately 560B tokens.

\subsection{Conditioning Fine-Tuning}
\label{conditioning-fine-tuning}
After pre-training and mid-training, the model was adapted in two shorter annealing stages of $10{,}000$ steps each. Both stages introduced the non-text conditioning mechanisms used at inference, removed datasets with undesirable characteristics, and increased inter-transcript agreement relative to mid-training. During training, the post-\ac{LDA} speaker embedding $\hat{\mathbf{e}}_\mathrm{x}$ for each utterance was passed through a learned projection and inserted before the text sequence. Because this embedding is high-bandwidth and carries information about the target audio, it can overwrite the same properties that the quality and speaking-rate conditioning are meant to control. The association between the embedding and these properties must be broken to ensure low-bandwidth conditioning is not overridden by the higher-bandwidth embedding.

This decorrelation was achieved by augmenting the source audio before the embedding was computed. The augmentations include environmental noise or music, frequency filters, audio compression artifacts, and reverberation. These were applied with a fixed probability $\alpha_\mathrm{AUG}$. To extend the training horizon of the speaker cloning stage we embedded a random crop of the target audio during the first annealing stage and masked the loss over the tokens in this crop. This reduces the incentive for the model to `cheat' using the shortcut, since it removes the loss on the specific segment of the audio the speaker embedding is computed on.

Quality conditioning is also introduced in the first annealing stage. Acoustic properties of the training audio, such as estimated \ac{SNR}, loudness, and signal bandwidth, are encoded as tokens using the same bucketing scheme as the speech rate. The speaking-rate and quality tokens are independently dropped with probabilities $0.4$ and $0.25$, respectively, so that the model learns both conditional and unconditional generation.

In the second annealing stage, the speaker embedding is computed from the entire target sequence rather than from a cropped portion of it. This ties the entire generated clip to the conditioning speaker and prevents the model from producing other speakers in the output. Because the embedding now spans the full clip, the loss masking described above is removed, as applying it would mask the entire clip and leave no frames to train on. Finally, a `Quality Mode' token is added when training on the highest-quality subset of the training data.

\begin{figure*}
    \centering

    \begin{subfigure}{0.48\textwidth}
        \centering
        \includegraphics[width=\linewidth]{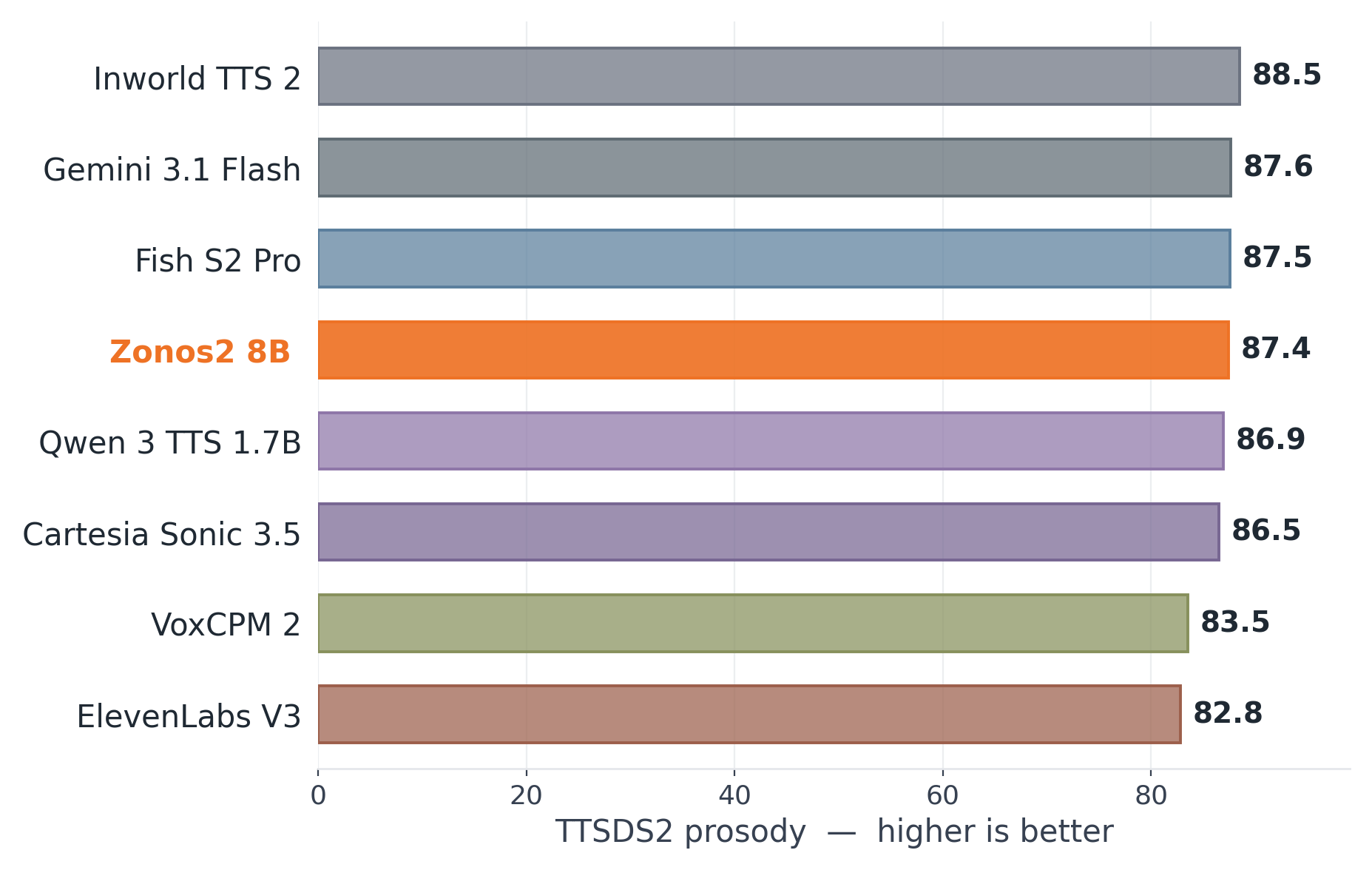}
        \caption{Clean set.}
        \label{fig:prosody-bar-clean}
    \end{subfigure}
    \hfill
    \begin{subfigure}{0.48\textwidth}
        \centering
        \includegraphics[width=\linewidth]{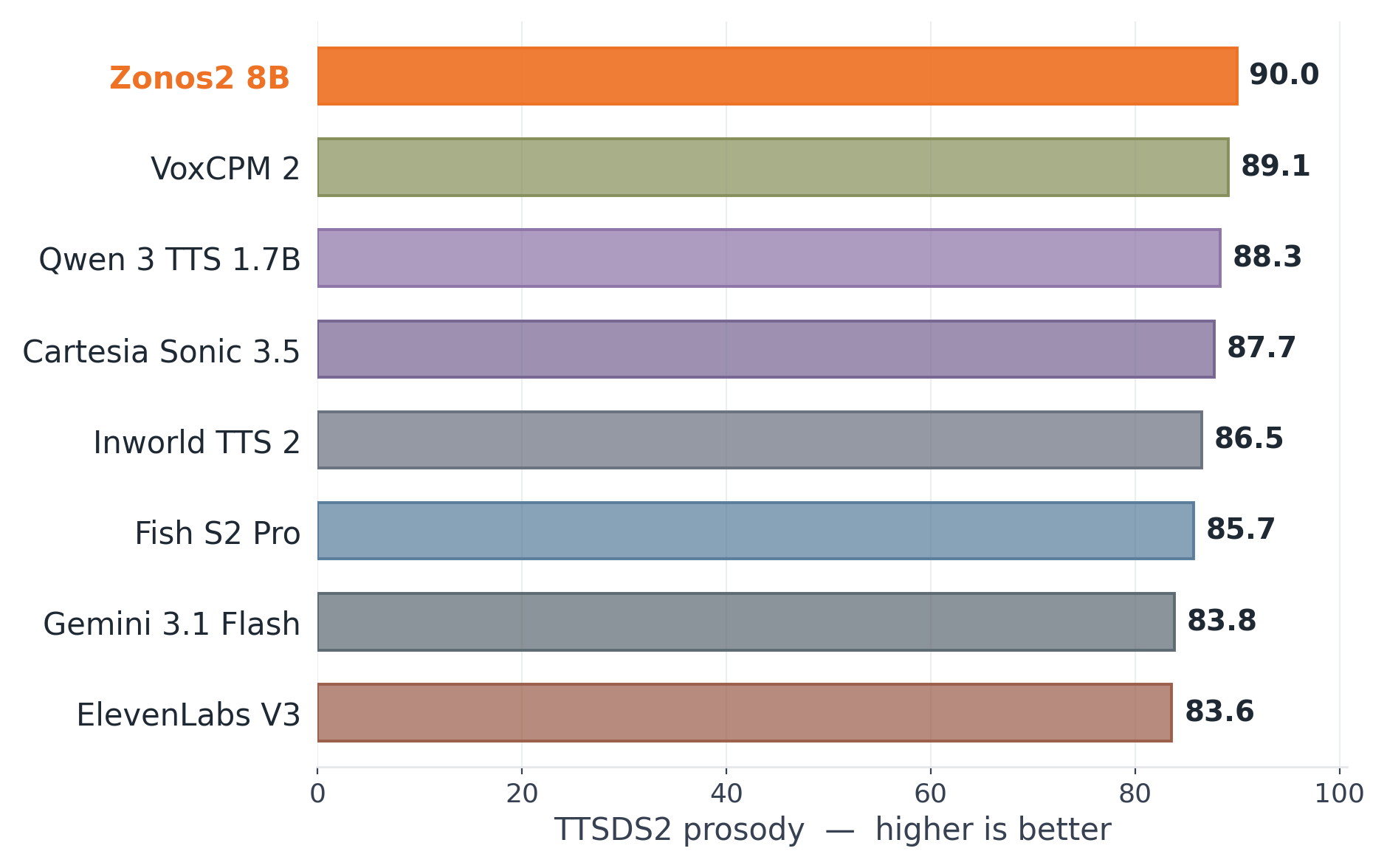}
        \caption{ITW set.}
        \label{fig:prosody-bar-itw}
    \end{subfigure}

    \caption{Mean \ac{TTSDS2} prosody for the English portions of both ZTTS1-Eval sets.}
    \label{fig:prosody-bars}
\end{figure*}

\begin{table*}[t]
\centering
\scriptsize
\setlength{\tabcolsep}{3.0pt}
\renewcommand{\arraystretch}{1.15}
\begin{tabular}{llrrrrrrrrrrr}
\toprule
\textbf{Model} & \textbf{Metric} & \textbf{en} & \textbf{hard\_en} & \textbf{zh} & \textbf{hard\_zh} & \textbf{de} & \textbf{es} & \textbf{fr} & \textbf{it} & \textbf{ja} & \textbf{ko} & \textbf{ru} \\
\midrule
\multicolumn{13}{l}{\textbf{Reference}} \\
\midrule
\multirow{3}{*}{\texttt{Ground truth}}
& WER $\downarrow$ & 3.53 & 1.04 & 2.83 & 1.28 & 4.68 & 3.23 & 3.03 & 5.09 & 4.06 & 2.88 & 6.12 \\
& UTMOS $\uparrow$ & 3.66 & 2.73 & 3.26 & 2.63 & 3.33 & 3.20 & 3.47 & 3.07 & 3.47 & 3.29 & 3.30 \\
& Spk. sim. $\uparrow$ & -- & -- & -- & -- & -- & -- & -- & -- & -- & -- & -- \\
\midrule
\midrule
\multicolumn{13}{l}{\textbf{Open-source models}} \\
\midrule
\multirow{3}{*}{\texttt{ZONOS2 8B }}
& WER $\downarrow$ & 2.76 & 15.04 & 15.62 & 26.23 & 5.67 & 4.78 & 13.18 & 6.53 & 8.27 & 18.96 & 8.26 \\
& UTMOS $\uparrow$ & 3.40 & 2.57 & 3.10 & 2.29 & 3.23 & 2.96 & 3.27 & 2.89 & 3.23 & 3.13 & 2.97 \\
& Spk. sim. $\uparrow$ & \underline{78.6} & 62.1 & 73.3 & 74.6 & 78.3 & 79.4 & 70.0 & 77.7 & 81.3 & 73.3 & 80.2 \\
\midrule
\multirow{3}{*}{\texttt{ZONOS2 8B Quality Mode}}
& WER $\downarrow$ & 3.99 & 2.68 & 6.73 & 14.41 & 3.74 & 3.25 & 4.30 & 3.52 & 7.67 & 4.14 & 6.99 \\
& UTMOS $\uparrow$ & 3.47 & 2.94 & 3.21 & 2.65 & 3.36 & 2.94 & 3.31 & 2.92 & 3.31 & 3.18 & 3.02 \\
& Spk. sim. $\uparrow$ & 74.4 & 58.2 & 81.1 & 73.4 & 76.2 & 79.0 & 75.9 & 78.0 & 82.0 & 83.0 & 79.4 \\
\midrule
\multirow{3}{*}{\shortstack[l]{\texttt{Qwen 3 TTS 1.7B}\\\citep{hu2026qwen3ttstechnicalreport}}}
& WER $\downarrow$ & \textbf{1.94} & 1.26 & \textbf{2.91} & 6.04 & \textbf{2.78} & \textbf{1.98} & \textbf{3.08} & \textbf{2.19} & \underline{3.78} & \textbf{2.85} & \textbf{4.25} \\
& UTMOS $\uparrow$ & \underline{3.86} & 3.33 & \textbf{3.71} & 3.16 & \textbf{3.72} & \textbf{3.43} & \textbf{3.64} & \underline{3.38} & \textbf{3.71} & \textbf{3.53} & \textbf{3.49} \\
& Spk. sim. $\uparrow$ & 68.3 & 60.2 & 79.7 & 75.8 & 69.7 & 69.2 & 72.0 & 75.2 & 81.0 & 79.3 & 74.1 \\
\midrule
\multirow{3}{*}{\shortstack[l]{\texttt{Fish S2 Pro}\\\citep{liao2026fishaudios2technical}}}
& WER $\downarrow$ & 3.60 & 5.00 & 4.33 & 7.95 & \underline{3.04} & 3.73 & 5.90 & 4.17 & 4.20 & 3.92 & 7.74 \\
& UTMOS $\uparrow$ & 3.47 & 3.06 & 3.23 & 3.00 & 3.26 & 3.01 & 3.22 & 3.02 & 3.37 & 3.18 & 3.09 \\
& Spk. sim. $\uparrow$ & 76.9 & 64.7 & \underline{82.9} & 76.0 & 78.5 & \underline{82.8} & \underline{78.0} & 79.3 & \underline{85.4} & \underline{83.9} & \underline{80.6} \\
\midrule
\multirow{3}{*}{\shortstack[l]{\texttt{VoxCPM 2}\\\citep{voxcpm2_2026}}}
& WER $\downarrow$ & 4.23 & 0.94 & 5.01 & \textbf{5.21} & 4.84 & 4.44 & 5.10 & 5.80 & 4.92 & 3.99 & 7.43 \\
& UTMOS $\uparrow$ & 3.51 & 2.82 & 3.12 & 2.55 & 3.07 & 2.85 & 3.25 & 2.71 & 3.23 & 3.04 & 2.83 \\
& Spk. sim. $\uparrow$ & 65.2 & \underline{66.8} & 77.3 & \underline{77.5} & \underline{79.7} & 75.4 & 75.8 & \underline{80.8} & 83.3 & 83.1 & 80.4 \\
\midrule
\multicolumn{13}{l}{\textbf{Closed-source models}} \\
\midrule
\multirow{3}{*}{\shortstack[l]{\texttt{Cartesia Sonic 3.5}\\\citep{cartesia_sonic35}}}
& WER $\downarrow$ & 2.56 & 0.86 & 4.53 & 7.66 & 3.14 & \underline{3.21} & \underline{3.41} & 3.05 & \textbf{3.65} & 3.24 & 6.01 \\
& UTMOS $\uparrow$ & 3.62 & 3.19 & 3.17 & 3.01 & 3.18 & 2.97 & 3.29 & 2.90 & 3.36 & 3.11 & 3.05 \\
& Spk. sim. $\uparrow$ & \textbf{79.9} & \textbf{67.1} & \textbf{85.4} & \textbf{80.0} & \textbf{82.8} & \textbf{85.9} & \textbf{83.1} & \textbf{83.4} & \textbf{87.5} & \textbf{86.4} & \textbf{84.2} \\
\midrule
\multirow{3}{*}{\shortstack[l]{\texttt{Eleven Labs V3}\\\citep{elevenlabs_v3}}}
& WER $\downarrow$ & \underline{2.35} & \textbf{0.75} & 4.20 & 8.31 & 3.79 & 3.24 & 3.62 & \underline{2.70} & 4.26 & 4.57 & \underline{5.63} \\
& UTMOS $\uparrow$ & 3.59 & \underline{3.53} & 3.31 & \underline{3.37} & 3.54 & 3.25 & 3.29 & 3.19 & 3.45 & 3.36 & 3.15 \\
& Spk. sim.* $\uparrow$ & 7.2 & 6.8 & 36.9 & 29.4 & 17.1 & 18.0 & 21.9 & 21.9 & 36.3 & 38.6 & 25.4 \\
\midrule
\multirow{3}{*}{\shortstack[l]{\texttt{Gemini 3.1 Flash}\\\citep{gemini31flash}}}
& WER $\downarrow$ & 2.50 & \underline{0.80} & 4.09 & \underline{5.63} & 3.61 & 4.41 & 4.77 & 3.24 & 3.79 & \underline{3.06} & 6.58 \\
& UTMOS $\uparrow$ & \textbf{3.87} & \textbf{3.79} & \underline{3.54} & \textbf{3.37} & \underline{3.71} & \underline{3.43} & \underline{3.52} & \textbf{3.39} & \underline{3.65} & \underline{3.51} & \underline{3.37} \\
& Spk. sim.* $\uparrow$ & 12.7 & 9.8 & 36.6 & 28.6 & 17.7 & 19.5 & 20.9 & 20.4 & 33.6 & 36.2 & 23.5 \\
\midrule
\multirow{3}{*}{\shortstack[l]{\texttt{Inworld TTS 2}\\\citep{inworld_realtime_tts2}}}
& WER $\downarrow$ & 3.14 & 1.96 & \underline{4.04} & 7.86 & 4.05 & 4.38 & 4.95 & 3.98 & 5.63 & 3.13 & 7.08 \\
& UTMOS $\uparrow$ & 3.53 & 3.15 & 3.14 & 2.94 & 3.27 & 3.05 & 3.33 & 2.95 & 3.27 & 3.11 & 3.01 \\
& Spk. sim. $\uparrow$ & 65.8 & 54.8 & 74.8 & 67.1 & 68.4 & 73.0 & 68.5 & 70.4 & 77.7 & 78.3 & 69.8 \\
\bottomrule
\end{tabular}
\caption{ZTTS1-Eval Clean zero-shot results across WER, MSR-UTMOS and speaker similarity, segmented by open-source and closed-source models. Best result per language and metric is shown in bold; second-best is underlined. Ground-truth reference results are shown for context and excluded from the ranking. * denotes models which do not support zero-shot voice cloning.}
\label{tab:clean-zero-shot-notext-merged}
\end{table*}

\begin{table*}[t]
\centering
\scriptsize
\setlength{\tabcolsep}{1.8pt}
\renewcommand{\arraystretch}{1.15}
\begin{tabular}{llrrrrrrrrrrrrrrrrr}
\toprule
\textbf{Model} & \textbf{Metric} & \textbf{en} & \textbf{zh} & \textbf{ar} & \textbf{de} & \textbf{es} & \textbf{fr} & \textbf{hi} & \textbf{id} & \textbf{it} & \textbf{ja} & \textbf{ko} & \textbf{pl} & \textbf{pt} & \textbf{ru} & \textbf{th} & \textbf{tl} & \textbf{tr} \\
\midrule
\multicolumn{19}{l}{\textbf{Reference}} \\
\midrule
\multirow{3}{*}{\texttt{Ground truth}}
& WER $\downarrow$ & 5.61 & 7.78 & 21.05 & 6.77 & 7.34 & 8.08 & 14.20 & 17.23 & 6.94 & 9.18 & 8.86 & 14.10 & 7.56 & 12.38 & 86.41 & 25.75 & 20.86 \\
& UTMOS $\uparrow$ & 2.48 & 2.39 & 2.22 & 2.52 & 2.22 & 2.27 & 2.22 & 2.14 & 2.19 & 2.27 & 2.48 & 2.41 & 2.23 & 2.23 & 2.40 & 2.24 & 2.27 \\
& Spk. sim. $\uparrow$ & 75.9 & 82.1 & 81.0 & 78.5 & 78.0 & 77.6 & 78.7 & 80.0 & 78.7 & 82.3 & 79.7 & 82.5 & 79.5 & 79.2 & 82.5 & 78.9 & 80.8 \\
\midrule
\midrule
\multicolumn{19}{l}{\textbf{Open-source models}} \\
\midrule
\multirow{3}{*}{\texttt{ZONOS2 8B }}
& WER $\downarrow$ & 4.70 & 3.19 & 21.43 & 5.84 & 5.38 & 4.56 & 15.50 & 11.84 & 5.61 & 10.18 & 7.45 & 10.16 & 6.73 & 7.80 & 12.87 & 23.49 & 10.46 \\
& UTMOS $\uparrow$ & 2.44 & 2.43 & 2.22 & 2.52 & 2.42 & 2.37 & 2.47 & 2.25 & 2.32 & 2.34 & 2.55 & 2.48 & 2.33 & 2.30 & 2.43 & 2.52 & 2.41 \\
& Spk. sim. $\uparrow$ & 67.0 & 74.3 & 67.4 & 69.4 & 69.4 & 67.7 & 66.3 & \underline{71.9} & 69.2 & 70.9 & 72.1 & 72.9 & 70.4 & 70.9 & 68.4 & \underline{68.5} & 72.1 \\
\midrule
\multirow{3}{*}{\texttt{ZONOS2 8B Quality Mode}}
& WER $\downarrow$ & 2.21 & 2.77 & 13.94 & 3.37 & 2.10 & 2.79 & 9.04 & 6.48 & 2.51 & 8.70 & 4.59 & 5.85 & 2.94 & 4.32 & 5.93 & \underline{16.05} & 7.47 \\
& UTMOS $\uparrow$ & 2.99 & 2.68 & 2.51 & 2.92 & 2.72 & 2.74 & 2.73 & 2.66 & 2.63 & 2.69 & 2.78 & 2.76 & 2.75 & 2.62 & 2.74 & 2.88 & 2.64 \\
& Spk. sim. $\uparrow$ & 56.9 & 70.6 & 63.8 & 63.1 & 63.3 & 61.2 & 62.3 & 67.6 & 63.8 & 70.0 & 67.3 & 69.0 & 63.8 & 67.7 & 68.3 & 65.0 & 67.7 \\
\midrule
\multirow{3}{*}{\shortstack[l]{\texttt{Qwen 3 TTS 1.7B}\\\citep{hu2026qwen3ttstechnicalreport}}}
& WER $\downarrow$ & \textbf{1.05} & \textbf{0.99} & -- & \textbf{2.11} & \underline{1.77} & 2.62 & -- & 12.46 & 2.85 & \underline{2.32} & \underline{2.80} & 82.04 & 2.20 & 3.88 & 9.61 & 18.04 & 83.20 \\
& UTMOS $\uparrow$ & 3.20 & 2.90 & -- & 3.12 & 2.87 & 2.84 & -- & \underline{2.94} & 2.76 & 2.86 & 2.97 & 2.92 & 2.85 & 2.74 & \underline{2.93} & \underline{3.13} & 2.70 \\
& Spk. sim. $\uparrow$ & 61.5 & 75.3 & -- & 68.3 & 65.8 & 67.1 & -- & 67.4 & 71.9 & 75.2 & 73.7 & 62.7 & 69.4 & 72.8 & 73.1 & 60.5 & 62.1 \\
\midrule
\multirow{3}{*}{\shortstack[l]{\texttt{Fish S2 Pro}\\\citep{liao2026fishaudios2technical}}}
& WER $\downarrow$ & 2.09 & 1.26 & 15.59 & 2.65 & 2.97 & 3.45 & 11.69 & 11.02 & 2.79 & \textbf{2.17} & 3.48 & 8.25 & 2.86 & 5.28 & 74.12 & 19.13 & 7.64 \\
& UTMOS $\uparrow$ & 2.92 & 2.73 & 2.58 & 2.90 & 2.69 & 2.67 & 2.69 & 2.72 & 2.54 & 2.73 & 2.76 & 2.67 & 2.57 & 2.56 & 2.75 & 2.89 & 2.58 \\
& Spk. sim. $\uparrow$ & 65.0 & 75.5 & \underline{72.0} & 69.4 & 68.2 & 68.3 & 69.9 & 71.4 & 70.4 & 75.0 & 72.8 & 73.5 & 71.8 & 72.4 & \underline{75.5} & 64.9 & 74.2 \\
\midrule
\multirow{3}{*}{\shortstack[l]{\texttt{VoxCPM 2}\\\citep{voxcpm2_2026}}}
& WER $\downarrow$ & 1.69 & 1.44 & \underline{12.18} & 2.84 & 3.19 & 4.54 & \textbf{7.13} & \underline{6.43} & 3.50 & 4.56 & 4.89 & 7.28 & 4.09 & 5.37 & \underline{4.74} & \textbf{15.70} & \underline{6.26} \\
& UTMOS $\uparrow$ & 2.51 & 2.40 & 2.25 & 2.60 & 2.39 & 2.36 & 2.41 & 2.30 & 2.26 & 2.30 & 2.51 & 2.35 & 2.32 & 2.30 & 2.39 & 2.51 & 2.32 \\
& Spk. sim. $\uparrow$ & \underline{68.1} & \underline{78.7} & \textbf{74.7} & \underline{72.8} & \underline{70.4} & \underline{72.2} & \underline{73.4} & \textbf{76.9} & \underline{74.5} & \underline{77.1} & \underline{74.8} & \underline{78.0} & \underline{73.5} & \underline{75.9} & \textbf{78.6} & \textbf{73.0} & \underline{76.9} \\
\midrule
\multicolumn{19}{l}{\textbf{Closed-source models}} \\
\midrule
\multirow{3}{*}{\shortstack[l]{\texttt{Cartesia Sonic 3.5}\\\citep{cartesia_sonic35}}}
& WER $\downarrow$ & 1.40 & \underline{1.17} & -- & 3.00 & 2.37 & \underline{2.52} & 7.87 & -- & 2.40 & 3.63 & 3.38 & \underline{5.30} & \underline{2.10} & \textbf{3.75} & -- & -- & \textbf{4.84} \\
& UTMOS $\uparrow$ & 3.05 & 2.76 & -- & 2.98 & 2.71 & 2.69 & 2.71 & -- & 2.60 & 2.69 & 2.86 & 2.73 & 2.68 & 2.64 & -- & -- & 2.57 \\
& Spk. sim. $\uparrow$ & \textbf{70.2} & \textbf{79.1} & -- & \textbf{76.7} & \textbf{75.4} & \textbf{75.5} & \textbf{76.1} & -- & \textbf{77.2} & \textbf{78.4} & \textbf{77.8} & \textbf{79.5} & \textbf{77.6} & \textbf{78.2} & -- & -- & \textbf{78.5} \\
\midrule
\multirow{3}{*}{\shortstack[l]{\texttt{Eleven Labs V3}\\\citep{elevenlabs_v3}}}
& WER $\downarrow$ & \underline{1.35} & 1.22 & 12.19 & 2.50 & \textbf{1.58} & \textbf{2.17} & \underline{7.36} & \textbf{5.98} & \underline{2.36} & 2.84 & 3.93 & \textbf{4.66} & \textbf{2.06} & 4.28 & \textbf{3.84} & 17.76 & 7.78 \\
& UTMOS $\uparrow$ & \underline{3.61} & \underline{3.36} & \textbf{3.41} & \underline{3.57} & \underline{3.34} & \underline{3.28} & \underline{3.47} & \textbf{3.34} & \underline{3.21} & \underline{3.49} & \underline{3.40} & \underline{3.37} & \underline{3.39} & \underline{3.26} & \textbf{3.48} & \textbf{3.51} & \underline{3.44} \\
& Spk. sim.* $\uparrow$ & 6.3 & 29.7 & 24.4 & 15.3 & 16.4 & 13.6 & 16.3 & 26.2 & 19.6 & 28.6 & 28.7 & 23.6 & 22.1 & 19.7 & 33.9 & 13.1 & 26.4 \\
\midrule
\multirow{3}{*}{\shortstack[l]{\texttt{Gemini 3.1 Flash}\\\citep{gemini31flash}}}
& WER $\downarrow$ & 2.12 & 1.45 & \textbf{10.51} & \underline{2.40} & 1.94 & 3.56 & 8.73 & -- & \textbf{2.08} & 2.72 & \textbf{2.27} & 5.33 & 3.21 & \underline{3.77} & -- & -- & 6.92 \\
& UTMOS $\uparrow$ & \textbf{3.78} & \textbf{3.55} & \underline{3.40} & \textbf{3.67} & \textbf{3.41} & \textbf{3.45} & \textbf{3.60} & -- & \textbf{3.45} & \textbf{3.56} & \textbf{3.45} & \textbf{3.47} & \textbf{3.43} & \textbf{3.36} & -- & -- & \textbf{3.66} \\
& Spk. sim.* $\uparrow$ & 10.2 & 29.9 & 26.3 & 17.5 & 20.5 & 15.0 & 18.1 & -- & 19.1 & 28.3 & 29.2 & 26.2 & 24.1 & 20.9 & -- & -- & 26.7 \\
\midrule
\multirow{3}{*}{\shortstack[l]{\texttt{Inworld TTS 2}\\\citep{inworld_realtime_tts2}}}
& WER $\downarrow$ & 1.89 & 1.42 & 14.85 & 3.54 & 2.67 & 4.40 & 10.14 & -- & 3.46 & 6.93 & 4.16 & 7.65 & 3.87 & 5.32 & -- & 19.56 & -- \\
& UTMOS $\uparrow$ & 3.02 & 2.71 & 2.70 & 3.03 & 2.83 & 2.76 & 2.65 & -- & 2.71 & 2.81 & 2.86 & 2.80 & 2.70 & 2.66 & -- & 2.99 & -- \\
& Spk. sim. $\uparrow$ & 53.0 & 66.4 & 60.5 & 58.7 & 56.2 & 54.9 & 60.5 & -- & 56.0 & 63.3 & 62.6 & 60.6 & 61.4 & 61.8 & -- & 59.5 & -- \\
\bottomrule
\end{tabular}
\caption{ZTTS1-Eval In-the-wild zero-shot results across WER, MSR-UTMOS, and speaker similarity segmented by open-source and closed-source models. Best result per language and metric is shown in bold; second-best is underlined. Ground-truth reference results are shown for context and excluded from the ranking. Note that some languages are unsupported by the baseline models. * denotes models which do not support zero-shot voice cloning.}
\label{tab:itw-zero-shot-notext-merged}
\end{table*}

\section{ZTTS1-Eval}\label{sect:ztts1}

Alongside the ZONOS2 model, we propose a new evaluation dataset and pipeline for expressive, voice-cloning-enabled \ac{TTS} systems, ZTTS1-Eval.

The design of ZTTS1-Eval is based on the widely used Seed-TTS-Eval~\citep{anastassiou2024seedttsfamilyhighqualityversatile}. We observe that Seed-TTS-Eval has three primary issues. First, it consists of only Chinese and English speech, limiting its usefulness for the assessment of multilingual systems. Second, it relies on outdated models to compute its assessment metrics such as Whisper Large~\citep{radford2022robustspeechrecognitionlargescale} and Paraformer~\citep{gao2023paraformerfastaccurateparallel} for \ac{WER} accuracy and WavLM~\citep{Chen_2022} for speaker similarity. Third, audio is sourced from Common Voice~\citep{ardila2020commonvoicemassivelymultilingualspeech} for English and DiDiSpeech~\citep{guo2021didispeechlargescalemandarin} for Chinese, both of which consist of read speech with varied recording environments which are generally not representative of the use-case of modern \ac{TTS} systems.

Two subsequent benchmarks each extend Seed-TTS-Eval along a single axis. CV3-Eval~\citep{du2025cosyvoice}, released with CosyVoice~3, broadens coverage to nine languages and adds task variety, with objective subsets for multilingual, cross-lingual, and emotion cloning and subjective subsets for expressive and accented speech. It nonetheless retains the Seed-TTS-Eval scoring stack (Whisper-Large and Paraformer for content, ERes2Net for speaker similarity, DNSMOS for quality) and consists of largely prepared speech. The MiniMax multilingual test set~\citep{minimax-speech} extends language coverage furthest, to 24 languages, but provides only two read Common Voice reference prompts per language and defers scoring to the Seed-TTS-Eval protocol, reporting \ac{WER} and speaker similarity alone. Neither benchmark updates the scoring models, includes spontaneous in-the-wild speech, or quantifies prosodic distribution or generation diversity.

ZTTS1-Eval is built to close each of these gaps. We extend coverage from the two languages of Seed-TTS-Eval to as many as $17$, pairing a clean read-speech set of $9$ languages with a spontaneous in-the-wild set of $17$ so that both prepared and conversational speech are represented. We replace the dated scoring stack end to end, scoring content with Qwen3-ASR, speaker similarity with ReDimNet, and quality with MSR-UTMOS. Finally, we add a prosodic-distribution and generation-diversity dimension that prior benchmarks lack, measured with TTSDS2 and \ac{DS-WED}. Table~\ref{tab:tts-eval-comparison} sets ZTTS1-Eval against Seed-TTS-Eval, CV3-Eval, and MiniMax-ML.

The ZTTS1-Eval `Clean' set totals $13$ hours, comprising $500$ utterances drawn from FLEURS-R~\citep{ma2024fleursrrestoredmultilingualspeech} for each of $9$ languages: English, Chinese, German, Spanish, French, Italian, Japanese, Korean, and Russian. The text in this set is structured `read-aloud' prepared speech; `hard' portions are also provided for particularly complex English and Chinese utterances. The `in-the-wild' (ITW) set comprises $1{,}618$ utterances from VoxBlink2~\citep{lin2024voxblink2100kspeakerrecognition} totaling roughly $3$ hours across $17$ languages; see Table~\ref{tab:itw-breakdown} for a full breakdown. The content in the ITW set is conversational and natural, representing `spontaneous' speech.

\ac{ASR} for \ac{WER} calculation is performed by the multilingual Qwen3-ASR~\citep{Qwen3-ASR} system.
We compute quality scores for the audio samples using MSR-\ac{UTMOS}~\citep{nishikawa2025multisamplingfrequencynaturalnessmosprediction}. Speaker similarity between the source clone audio and the generated audio is computed using ReDimNet~\citep{Yakovlev_2024}. \ac{DS-WED}~\citep{yang2026dswed} is used to assess prosodic variation between pairs of generations from the same source clone audio, and TTSDS2-Prosody~\citep{minixhofer2025ttsds2} is used to assess general prosody.

Note that ZONOS2 is deliberately not trained on any of the audio in ZTTS1-Eval. We cannot speak to whether any of the comparable models without public training sets are trained on data used by ZTTS1-Eval.
ZTTS1-Eval is available on GitHub\footnote{https://github.com/Zyphra/ZTTS1-Eval}.

\section{Results}\label{sect:results}

Tables~\ref{tab:clean-zero-shot-notext-merged} and~\ref{tab:itw-zero-shot-notext-merged} show results for the ZTTS1-Eval sets for ZONOS2 as well as a number of open-source and closed-source baselines. Results are presented for ZONOS2 with and without the `Quality Mode' token set.
For both sets, the `Ground Truth' row shows the performance of the \ac{ASR} system on the clone target audio; for the ITW set, speaker similarity represents the similarity between two utterances from the same speaker.

On the Clean subset, shown in Table~\ref{tab:clean-zero-shot-notext-merged}, ZONOS2 shows competitive performance in terms of cloning accuracy as measured by speaker similarity, \textbf{achieving the best open-source and the second best overall speaker similarity for English~(en)}. The gain in intelligibility here, as measured by \ac{WER} via use of the `Quality Mode' token, is somewhat uneven. We observe degraded \ac{WER} scores for English but improvements for all other languages. Quality Mode drastically improves intelligibility for some languages such as Mandarin (zh) from $15.62\%$ to $6.73\%$, while also improving speaker similarity. Quality Mode also improves the acoustic quality of the generated audio as measured by \ac{UTMOS}.

On the ITW subset, shown in Table~\ref{tab:itw-zero-shot-notext-merged}, ZONOS2 performs well in terms of \ac{WER} and speaker similarity; here the positive effect of Quality Mode is most apparent, where it shows improved \ac{WER} and \ac{UTMOS} across all languages versus the base setting. The use of Quality Mode has a negative effect on speaker similarity but raises intelligibility and quality metrics.

\begin{figure*}[t]
    \centering
    \includegraphics[width=0.9\linewidth]{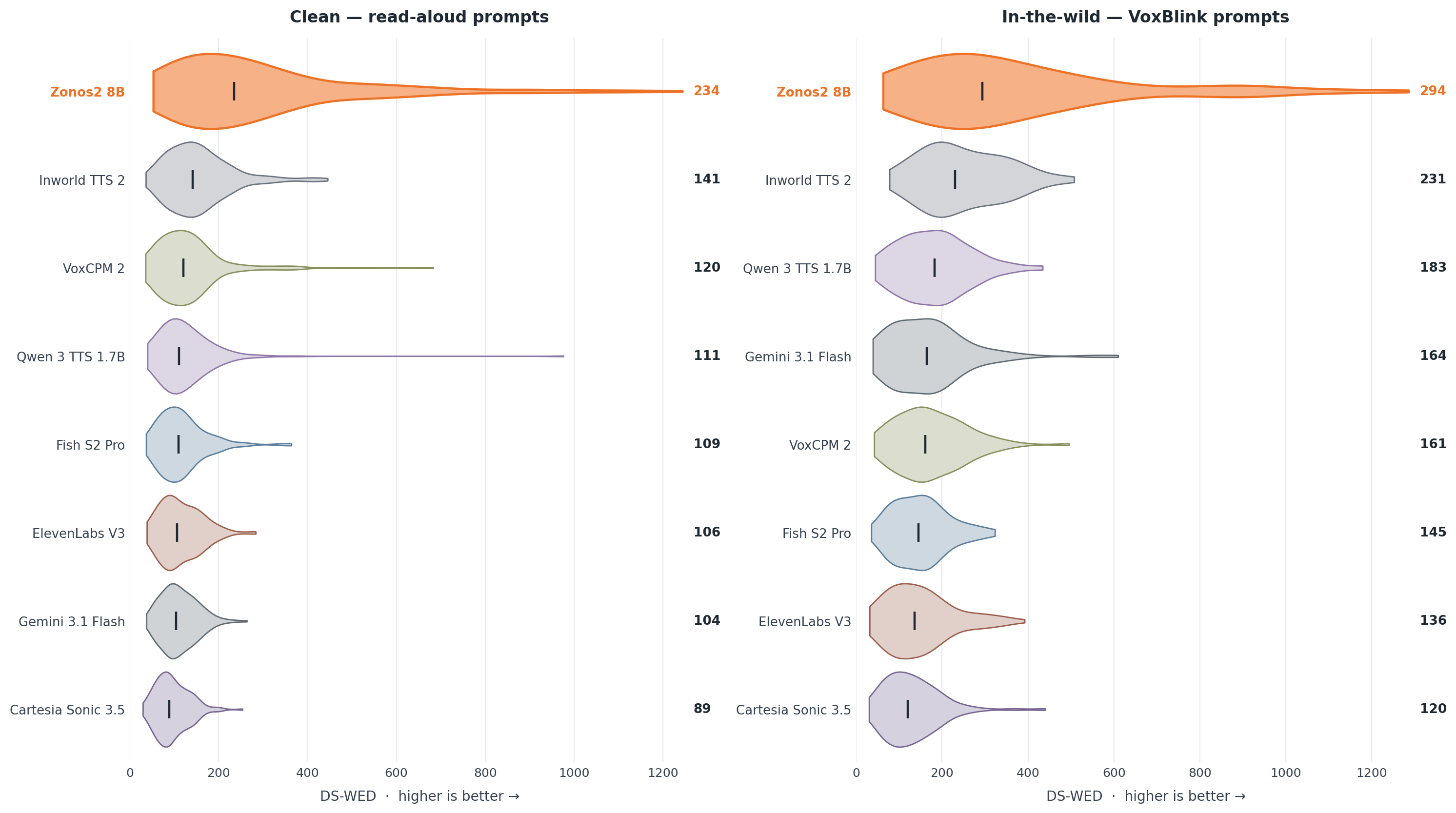}
    \caption{Violin plots of DS-WED scores for the English portions of both ZTTS1-Eval sets.}
    \label{fig:DS-WED-violins}
\end{figure*}

\begin{figure*}[t]
    \centering
    \includegraphics[width=0.9\linewidth]{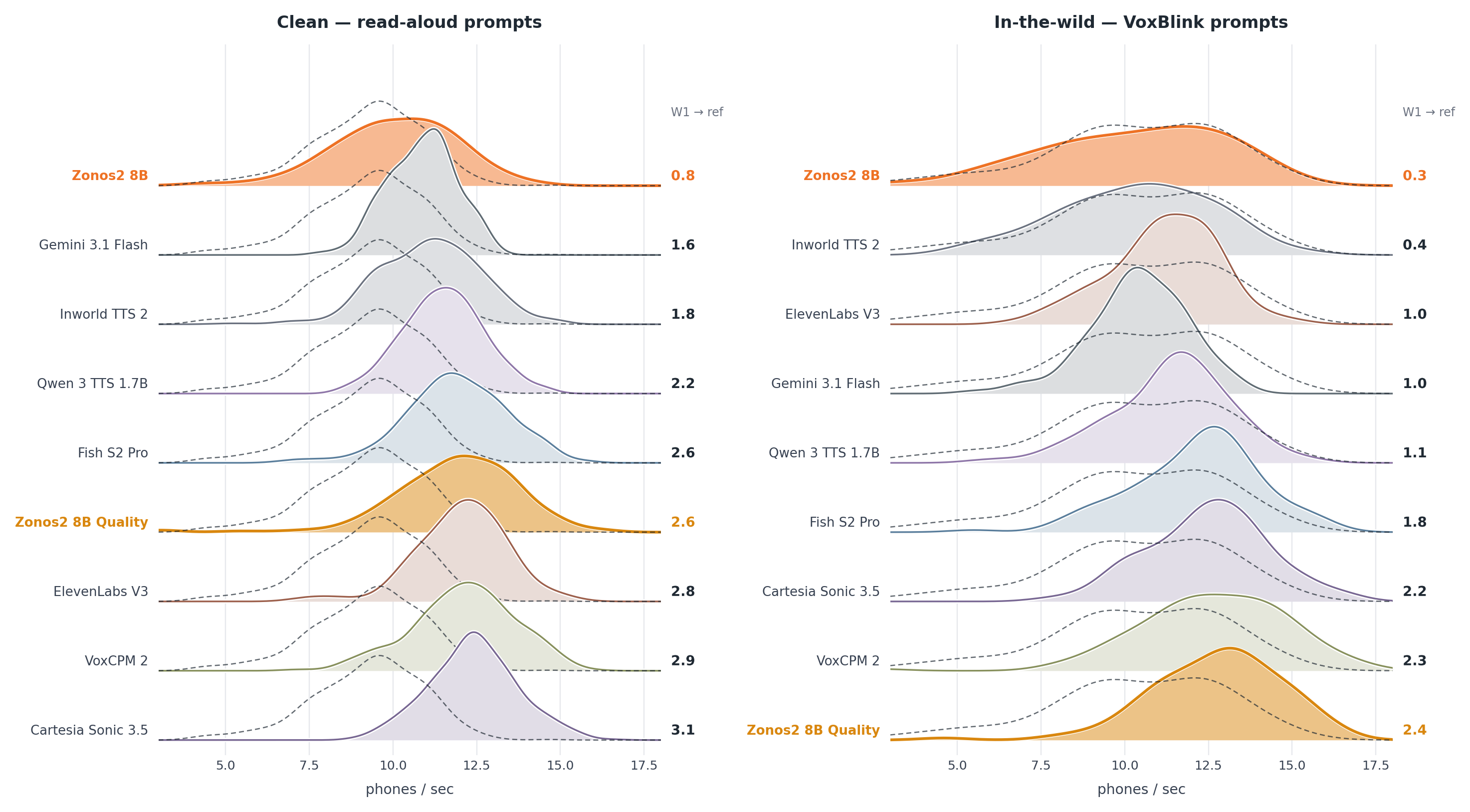}
    \caption{Allosaurus SR distributions for the English portions of both ZTTS1-Eval sets.}
    \label{fig:allosaurus_dist}
\end{figure*}

Figures~\ref{fig:prosody-bar-clean} and~\ref{fig:prosody-bar-itw} show the mean \ac{TTSDS2} prosody for the English portions of the Clean and ITW eval sets, respectively. On the ITW set, ZONOS2 is the best performing model for this metric. The performance of ZONOS2 on the Clean set is weaker, but remains competitive with the top performing models. These results demonstrate the ability of ZONOS2 to retain prosody information as well as identity from the source clone audio. 
Figure~\ref{fig:DS-WED-violins} further illustrates this, depicting the \ac{DS-WED} scores for the English subsets of ZTTS1-Eval. ZONOS2 demonstrates significantly higher prosodic variation in its generations relative to all other models across both eval sets. Figure~\ref{fig:allosaurus_dist} shows the distribution of the prosodic content relative to the source audio measured as Allosaurus~\citep{li2020universal} SR distance; ZONOS2 has a clear advantage here, showing a much closer distribution to that of the source.
See Table~\ref{tab:zonos2-other-eval} for ZONOS2 performance on the CosyVoice 3 and Seed-TTS evaluation sets.

\section{Discussion}\label{sect:discuss}
\subsection{Attention-Mechanism Selection}
During early ablations with both fully dense and \ac{MoE} backbones, it was found that \ac{MHA} was significantly more stable than \ac{GQA} for our data and architecture, and produced higher-quality outputs. However, in the interest of inference speed, we selected \ac{GQA}. We believe aligning high frame rate audio with low frame rate text has different demands on attention compared to next token prediction for text. Future work will explore the relative importance and function of attention in high frame rate \ac{TTS} models and how it compares to LLMs.

Experiments on all Qwen gating~\citep{qwen-gating} variants showed headwise gating to be most effective, with minimal impact on training or inference overhead.

\subsection{\ac{MoE} Balancing Problems and Solutions}
The integration of \ac{DAC} token prediction with the delay pattern in an \ac{MoE} backbone proved particularly challenging. When comparing to large language models of similar sizes with similar hyperparameters, balancing on delayed \ac{DAC} tokens was significantly more unstable than on pure text. The configuration of $3$ initial and $1$ final dense layers was found to be the most stable during initial testing, when combined with top-$2$ selection for the final \ac{MoE} layer. A large portion of the instability in the final ZONOS2 model can be attributed to the inherent difficulty of delayed \ac{DAC} token prediction. Future work will explore the use of alternative audio codecs to stabilize training, improve generation robustness and inference efficiency.

\subsection{Speaker Embedding Overfitting}
During implementation of the speaker cloning conditioning, we observed causal leakage of information regarding the length, lexical content, and pause distribution of the source clone audio which caused inference instability characterized by either silent output or `glossolalia' babble output. Several interventions were made to attempt to mitigate this, including an inference time regression probe of the embedding, as well as a number of training time changes. The most effective of these was the \ac{LDA} dimensionality reduction in combination with a two-stage anneal used in the final implementation. 

\section{Conclusion}
This work presents ZONOS2, a state-of-the-art \ac{TTS} system based on an \ac{MoE} transformer backbone. A complementary new benchmark for the assessment of similar systems, ZTTS1-Eval, is presented. On ZTTS1-Eval and other standard open-source TTS evaluation benchmarks, ZONOS2 shows comparable performance to both open- and closed-source systems and exceptional voice cloning capability. Future work will investigate the use of different audio codecs, backbone designs, and post-training strategies.

\section*{Acknowledgements}
We would like to thank Rishi Iyer for the discussions on MoE implementation and training dynamics, Nathan Kolbas for assistance in setting up and maintaining the ZONOS2 model endpoint on Zyphra Cloud, and Paul White and Danny Martinelli for assistance with the release and distribution of the ZONOS2 model.

\FloatBarrier
\bibliographystyle{tmlr}
\bibliography{main}

\appendices
\section{Model configuration}

\begin{table}[h]
\begin{tabular}{ll}
\hline
\textbf{Property}       & \textbf{ZONOS2 8B Configuration}               \\ \hline
Architecture            & Decoder-only MoE transformer \\
Active Parameters       & 900M                                            \\
Total Parameters        & 8B                                              \\
Transformer Layers      & 28                                              \\
Hidden Dimension        & 2048                                            \\
GQA Query Heads         & 16                                              \\
KV Heads                & 4                                               \\
Head dimension          & 128                                             \\
Attention variant       & GQA (4$\times$ grouping)                        \\                                 
Experts per MoE layer   & 16                                              \\
Routing                 & top-1, top-2 for final MoE                      \\
Expert FFN width        & 3072                                            \\
Qwen gating location & headwise \\
Router latent dimension & 128 \\
Router configuration & \ac{EDA} \\
Positional embeddings   & RoPE                                            \\
Tokenizer               & Byte level                                      \\
\hline
\end{tabular}
\caption{ZONOS2 8B Configuration.}\label{tab:zonos_config}
\end{table}

\section{ITW set language breakdown}

\begin{table}[h]
\centering
\small
\setlength{\tabcolsep}{8pt}
\renewcommand{\arraystretch}{1.15}
\begin{tabular}{lrr}
\toprule
\textbf{Language} & \textbf{Utterances} & \textbf{Hours} \\
\midrule
en & 120 & 0.22 \\
zh & 92 & 0.14 \\
de & 104 & 0.20 \\
es & 95 & 0.17 \\
fr & 92 & 0.18 \\
it & 97 & 0.18 \\
ja & 91 & 0.14 \\
ko & 93 & 0.13 \\
ru & 101 & 0.19 \\
pt & 93 & 0.19 \\
ar & 90 & 0.15 \\
hi & 86 & 0.14 \\
id & 84 & 0.16 \\
tr & 93 & 0.17 \\
tl & 89 & 0.14 \\
pl & 105 & 0.21 \\
th & 93 & 0.14 \\
\midrule
\textbf{Total} & \textbf{1618} & \textbf{2.86} \\
\bottomrule
\end{tabular}
\caption{ZTTS1-Eval ITW language coverage statistics.}
\label{tab:itw-breakdown}
\end{table}

\section{Model Details}\label{sect:model_details}
Each transformer layer follows a pre-normalization residual structure. For layer
$\ell$, with input hidden state $h^\ell_t$, the following is computed:
\begin{align}
    \bar{h}^\ell_t &= \mathrm{RMSNorm}_{\mathrm{attn}}(h^\ell_t), \\
    r^\ell_t &= h^\ell_t + \mathrm{Attn}^\ell(\bar{h}^\ell_{\le t}), \\
    h^{\ell+1}_t &= r^\ell_t +
    \mathrm{FFN}^\ell\!\left(\mathrm{RMSNorm}_{\mathrm{ffn}}(r^\ell_t)\right).
\end{align}
The attention module uses \ac{GQA}. Queries are projected into
$H$ query heads, while keys and values are projected into $H_{\mathrm{kv}}$
heads, with $H_{\mathrm{kv}} \le H$. \ac{RoPE}~\citep{SU2024127063} are applied to
queries and keys before attention. Query-key normalization is applied~\citep{henry-etal-2020-query}:
\begin{align}
    q_{t,h} &\leftarrow
    \alpha_h \frac{q_{t,h}}{\mathrm{RMS}(q_{t,h})}, \\
    k_{t,h} &\leftarrow
    \frac{k_{t,h}}{\mathrm{RMS}(k_{t,h})},
\end{align}
where $\alpha_h$ is a learned positive per-head scale. Attention is then
computed using FlashAttention~\citep{dao2022flashattentionfastmemoryefficientexact} over packed variable-length sequences:

\begin{equation}
    o_{t,h}
    =
    \sum_{\tau \le t}
    \operatorname{softmax}_{\tau}
    \left(
        \frac{q_{t,h}^{\top} k_{\tau,h}}{\sqrt{d_h}}
    \right)
    v_{\tau,h}.
\end{equation}
The attention output is modulated with a learned
head-wise gate~\citep{qwen-gating}:
\begin{equation}
    g_t = \sigma(W_g \bar{h}^\ell_t), \qquad
    \tilde{o}_{t,h} = g_{t,h} o_{t,h}.
\end{equation}
The gated attention output is then projected back to the model dimension.\\
The feed-forward sublayer is either a dense SwiGLU \citep{shazeer2020gluvariantsimprovetransformer} \ac{MLP} in the initial and final non-\ac{MoE} blocks or the routed
\ac{MoE} \ac{MLP} shown in the center of Figure~\ref{fig:moe_transformer}. The dense form is
\begin{equation}
    \mathrm{FFN}(x)
    =
    W_{\mathrm{out}}
    \left(
        W_{\mathrm{in},1}x \odot
        \operatorname{SiLU}(W_{\mathrm{in},2}x)
    \right).
\end{equation}
For routed layers, a router produces expert probabilities
\begin{equation}
    \pi_t = \operatorname{softmax}(R(x_t)),
\end{equation}
selects the top-$k$ experts and combines their outputs, weighted by the corresponding router probabilities. Routed layers use $E=16$ experts with top-$1$ routing, except the last \ac{MoE} block, which uses top-$2$. The \ac{MoE} layers begin after the first $3$ dense layers, and the final layer is also dense.

\begin{table*}[h]
\centering
\scriptsize
\setlength{\tabcolsep}{4.0pt}
\renewcommand{\arraystretch}{1.15}
\begin{tabular}{llrrrr}
\toprule
\textbf{Task} & \textbf{Set} & \textbf{Spk. sim. $\uparrow$} & \textbf{DNSMOS $\uparrow$} & \textbf{WER \% $\downarrow$} & \textbf{Emo. acc. \% $\uparrow$} \\
\midrule
\multicolumn{6}{l}{\texttt{ZONOS2}} \\
\midrule
\multicolumn{6}{l}{\textit{CosyVoice 3 Eval}} \\
\midrule
\multirow{6}{*}{\textbf{Zero-shot}} & en & 49.66 & 3.901 & 4.48 & -- \\
 & hard\_en & 49.58 & 4.020 & 5.23 & -- \\
 & zh & 56.93 & 3.710 & 12.08 & -- \\
 & hard\_zh & 50.64 & 3.614 & 25.55 & -- \\
 & ja & 53.15 & 3.846 & 8.60 & -- \\
 & ko & 58.75 & 3.945 & 6.03 & -- \\
\midrule
\multirow{6}{*}{\textbf{Cross-lingual zero-shot}} & to\_en & 46.66 & 3.896 & 4.94 & -- \\
 & to\_hard\_en & 45.93 & 4.007 & 9.13 & -- \\
 & to\_zh & 47.00 & 3.701 & 18.72 & -- \\
 & to\_hard\_zh & 47.60 & 3.665 & 23.27 & -- \\
 & to\_ja & 48.13 & 3.866 & 15.12 & -- \\
 & to\_ko & 50.40 & 3.918 & 6.82 & -- \\
\midrule
\multirow{2}{*}{\textbf{Emotion zero-shot}} & en & 39.12 & 3.837 & 4.71 & 37.3 \\
 & zh & 58.48 & 3.610 & 7.57 & 36.7 \\
\midrule
\multirow{4}{*}{\textbf{Subjective continuation}} & emotion & 48.16 & 3.743 & -- & -- \\
 & rhyme & 42.00 & 3.547 & -- & -- \\
 & speed & 45.25 & 3.768 & -- & -- \\
 & volume & 44.54 & 3.777 & -- & -- \\
\midrule
\multirow{1}{*}{\textbf{Subjective zero-shot}} & all & 53.32 & 3.791 & -- & -- \\
\midrule
\multicolumn{6}{l}{\textit{Seed-TTS-Eval}} \\
\midrule
\multirow{3}{*}{\textbf{Zero-shot}} & Test-EN & 47.60 & -- & 2.05 & -- \\
 & Test-ZH & 58.20 & -- & 2.55 & -- \\
 & Test-ZH-Hard & 56.2 & -- & 11.15 & -- \\
\bottomrule
\end{tabular}
\caption{ZONOS2 evaluation results across speaker similarity, DNSMOS, WER, and emotion accuracy on the CosyVoice 3 Eval and Seed-TTS-Eval benchmarks, grouped by task. Emotion accuracy is only reported for the emotion zero-shot task.}
\label{tab:zonos2-other-eval}
\end{table*}

\section{Additional Result Plots}
This appendix collects the supplementary prosody and generation-diversity plots referenced in Section~\ref{sect:results}.

\end{document}